\documentclass[aps,prl,showpacs,twocolumn,floatfix]{revtex4-1}
\usepackage{graphicx}
\usepackage{amssymb}
\usepackage{subfigure}

\usepackage{color} 
\usepackage{xcolor}

\begin{document}

\title{Atom radio-frequency interferometry}
\author{D.~A.~Anderson}
\email{dave@rydbergtechnologies.com}
\author{R.~E.~Sapiro}
\author{L.~F.~Gon\c{c}alves}
\author{R.~Cardman}
\author{G.~Raithel}
\affiliation{Rydberg Technologies Inc., Ann Arbor, MI 48103}

\date{\today }
\begin{abstract}
We realize and model a Rydberg-state atom interferometer for measurement of phase and intensity of radio-frequency (RF) electromagnetic waves. A phase reference is supplied to the atoms via a modulated laser beam, enabling atomic measurement of the RF wave's phase without an external RF reference wave.  The RF and optical fields give rise to closed interferometric loops within the atoms' internal Hilbert space. In our experiment, we construct interferometric loops in the state space $\{ 6P_{3/2}, 90S_{1/2}, 91S_{1/2}, 90P_{3/2} \}$ of cesium and employ them to measure phase and intensity of a 5~GHz RF wave in a room-temperature vapor cell. Electromagnetically induced transparency on the $6S_{1/2}$ to $6P_{3/2}$ transition serves as an all-optical interferometer probe. The RF phase is measured over a range of $\pi$, and a sensitivity of 2~mrad is achieved.  RF phase and amplitude measurements at sub-millimeter optical spatial resolution are demonstrated.
\end{abstract}

\maketitle

Interferometry has been foundational to the advancement of science and technology since its inception at the end of the 19th century~\cite{Michelson.1881}.  As a means to detect the phase of electromagnetic waves, interferometry is ubiquitous across disciplines from optics~\cite{born2013principles} and metrology~\cite{Chang.1965}, to cosmology~\cite{WolfQuantumLimOpticalInterferometry.2015,Spitler.2016}, biomedical imaging~\cite{Park.2018}, and optical~\cite{Gagliardi.1976} and radio-frequency (RF) communications~\cite{Flemming.1919}.  Wave-particle duality~\cite{DeBroglie.1924} and interferometry~\cite{Ramsey.1950} are fundamentally important
for quantum mechanics, atomic and molecular physics, and precision metrology~\cite{Cronin.2009}, from atom interferometry~\cite{Keith.1991,kasevich.1991} and establishing fundamental principles of quantum measurement~\cite{Brune.1996} to the emergence of quantum technologies~\cite{MacFarlane.2003} including atomic clocks~\cite{Essen.1955,Ludlow.2015} and inertial sensors~\cite{Wang.2005} for precision timing and navigation.

Interferometric phase sensing is often performed by linear superposition of electromagnetic signal and reference waves of the same or similar frequencies, and using respective homodyne or heterodyne detectors.  Multi-spectral phase-sensing of fields that spread across the electromagnetic spectrum, from DC to optical, can be accomplished by wave mixing in macroscopic nonlinear devices, such as nonlinear crystals, mixers, modulators and demodulators. At a fundamental level, phase-coherent, nonlinear wave mixing also occurs in an individual atom, whose Hilbert space provides a grid of states that can be connected via electromagnetic couplings ranging from DC to optical and beyond. In this work, we present an atom interferometer for RF electromagnetic waves: an interferometer in which optical fields and a single RF signal wave connect a pair of atomic states via two paths. The atom RF interferometer is employed for optical measurement of phase and intensity of an RF signal wave via readout of atomic populations~\cite{Anderson.2019,AndersonIEEEAES.2020}.

In our atom RF interferometer, RF phase-reference information is transmitted to the atoms via a RF-modulated optical carrier field, preparing the atomic medium in a coherent superposition state of matter that is sensitive to phase and intensity of an incident signal RF wave.  The interferometer is implemented with Rydberg states of atoms~\cite{Bethe.1957,Gallagher}, which are highly susceptible to electric fields ranging from DC to THz. As a readout for the interferometer we use Rydberg electromagnetically induced transparency (EIT)~\cite{Mohapatra.2007}, a method suitable for atom-based sensing in room-temperature vapors.  We employ the atom RF interferometer to optically receive and measure the phase and intensity of an RF signal wave.

\begin{figure}[h]
\includegraphics[width=6.5cm]{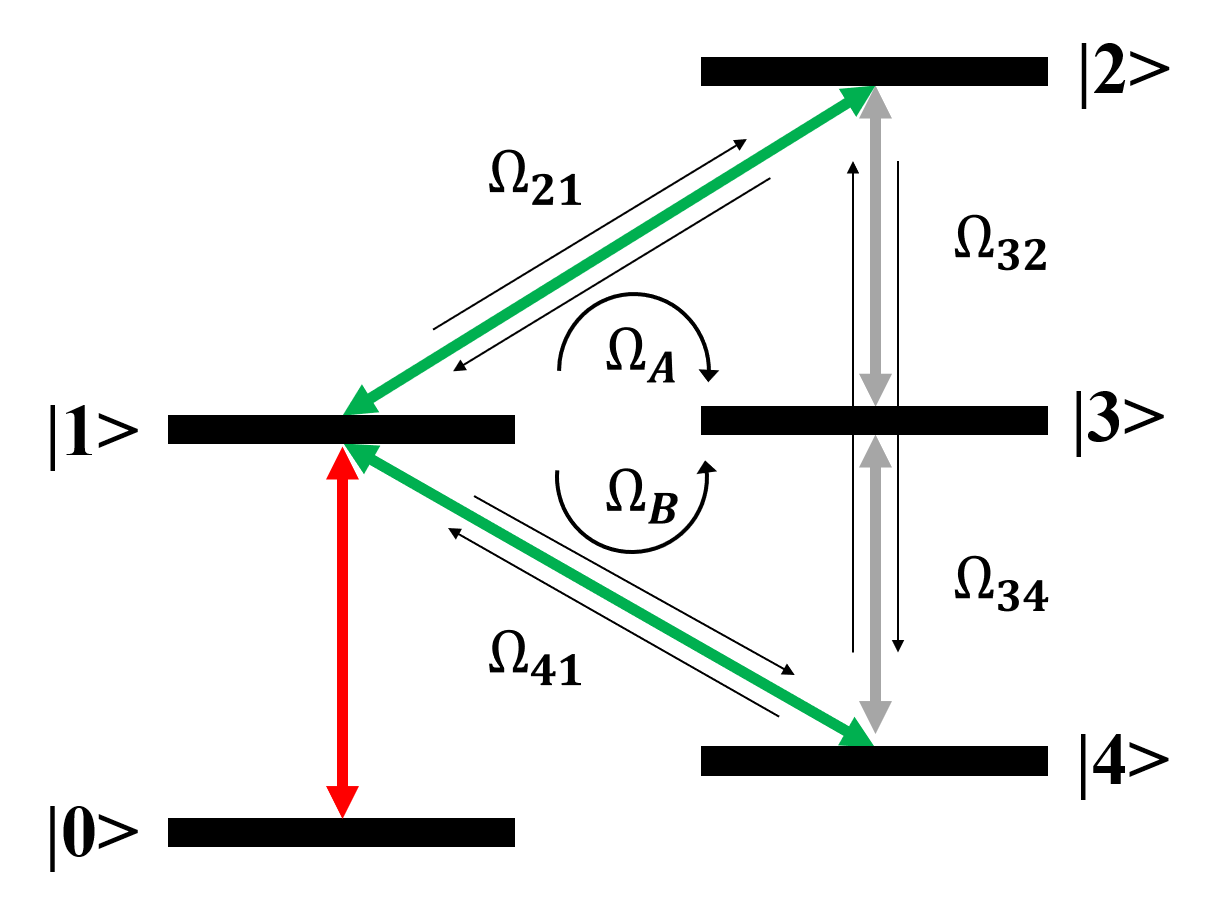}
\caption{Atom radio-frequency interferometer.  Atomic states $\vert2\rangle$ and $\vert3\rangle$ and states $\vert3\rangle$ and $\vert4\rangle$ are electric-dipole coupled by the same RF field (gray arrows); atomic states $\vert1\rangle$ and $\vert2\rangle$ and states $\vert1\rangle$ and $\vert4\rangle$ are coupled by two distinct optical fields (green arrows). All fields are phase-coherent. The interferometric pathways, indicated by the thin circular arrows, are closed within the quantum state-space.  An additional optical field resonant with the $\vert0\rangle$ to $\vert1\rangle$ transition (red arrow) provides an optical readout for the interferometer.}
\label{fig:1}
\end{figure}

RF sensing with Rydberg atoms employing fiducial RF reference waves to achieve phase-sensitive and field-enhanced RF sensing has been previously demonstrated~\cite{Anderson.2019, AndersonIEEEAES.2020,SimonsPhase.2019,Jing.2020}.  Our optical interferometric RF phase sensing approach realized in this work differs fundamentally from previous work~\cite{Sedlacek.2012,Holloway.2014,Anderson.2014,Anderson.2016,
Anderson2.2017,AndersonPlasma.2017,Meyerpub.2018,SimonsPhase.2019, Meyer2020,Jing.2020} in which atoms serve as sensors for the intensity of an RF wave. There, information on the relative phase between an RF signal and RF reference wave is extracted from atom-based RF field intensity measurements of the superposition of the external RF waves. In the present paper, we prepare and employ a coherent atomic medium that is sensitive to both phase and intensity of an RF signal wave, without requiring an external RF reference wave.

\begin{figure*}[t]
\includegraphics[width=18cm]{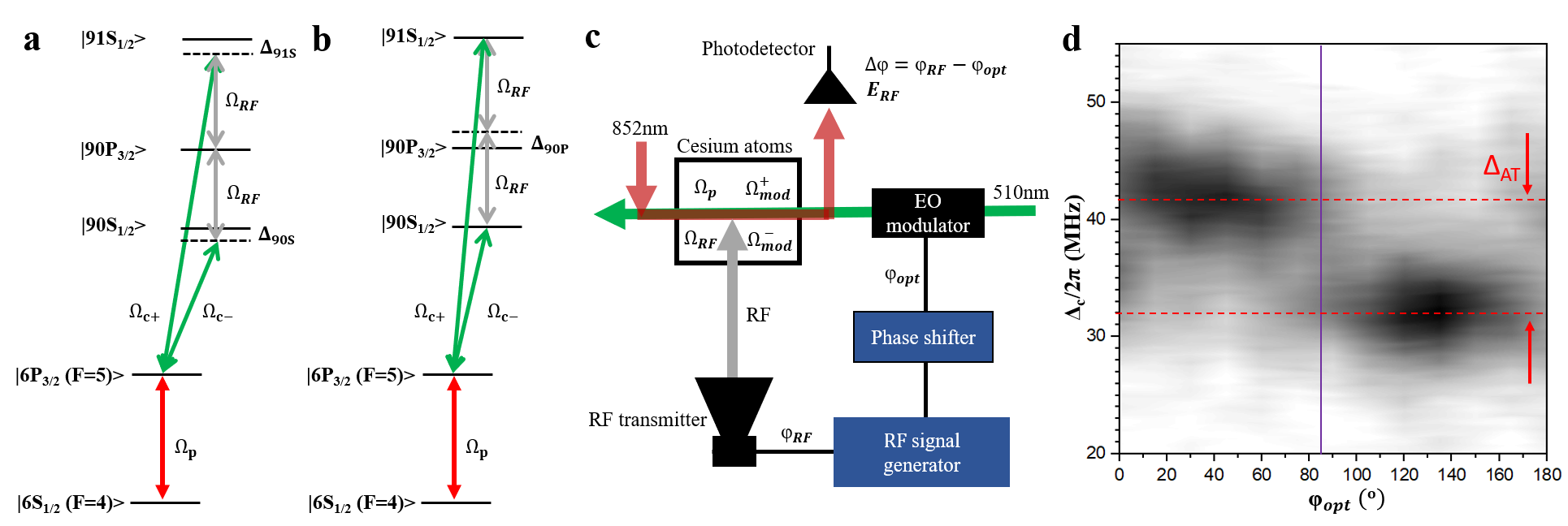}
\caption{(a) and (b) Energy-level diagrams for interferometric loops I and II in a cesium vapor with Rydberg EIT. (c) Experimental setup.  (d) Phase measurement of a 5.092~GHz electromagnetic field using interferometric loop II.}
\label{fig:2}
\end{figure*}

The atomic energy-level and coupling scheme~\cite{Anderson.2019,AndersonIEEEAES.2020} is illustrated in Fig.~\ref{fig:1}. There, $|0\rangle\rightarrow|1\rangle$ is an optical transition to probe the quantum interference. The $|1\rangle\rightarrow|2\rangle$ and $|1\rangle\rightarrow|4\rangle$ transitions are driven by optical fields (respective frequencies $\omega_{21}$ and $\omega_{41}$), and the $|2\rangle\rightarrow|3\rangle$ and $|4\rangle\rightarrow|3\rangle$ transitions by the signal RF field (frequency $\omega_{RF}$).  The frequency difference between the optical fields, $\omega_{41}-\omega_{21}$, equals $2\omega_{RF}$.  In this way, a closed pair of interfering pathways from $|1\rangle$ to $|3\rangle$ is established, labeled $A$ and $B$ in Fig.~\ref{fig:1}, with each path involving one optical and one RF photon.
The respective transition amplitudes are
\begin{eqnarray}
 \Omega_{A} &=& \frac{\Omega_{21} \, \Omega_{32}}{2 \Delta_{A}}
              \exp( {\rm{i}}(  \phi_{opt}-\phi_{RF}))  \nonumber \\
 \Omega_{B} &=& \frac{\Omega_{41} \, \Omega_{34}}{ 2 \Delta_{B}}
               \exp( {\rm{i}}(-\phi_{opt}+\phi_{RF}))
\label{eq:int1}
\end{eqnarray}
where the $\Omega_{nm}$ are (real-valued) magnitudes of the Rabi frequencies of the transitions from $|m\rangle$ to $|n\rangle$, and $\Delta_{A}$ and $\Delta_{B}$ are detunings that occur in the respective paths, as specified below.
The RF signal field carries a phase $\phi_{RF}$, and the pair of optical fields carries phases $ \pm \phi_{opt}$.

The closed interferometric loop exhibits quantum interference between the two pathways associated with $\Omega_{A}$ and $\Omega_{B}$ (see Fig.~1). In the case studied, we may set $\vert \Omega_A \vert = \vert \Omega_B \vert = \Omega_0$.
Phases accumulated along the optical beam paths are compounded in an optical phase difference $2 \phi_{opt}$. The  interferometric sum of the excitation amplitudes from $|1\rangle$ to $|3\rangle$ exhibits the atom-interferometric phase dependence,
\begin{equation}
\Omega = 2 \Omega_{0} \cos(\phi_{RF} - \phi_{opt}) \quad.
\label{eq:int2}
\end{equation}

Figures~\ref{fig:2}a and b show two implementations of the interferometric principle using Rydberg EIT~\cite{Mohapatra.2007} in a cesium room-temperature vapor cell. The 852-nm EIT probe laser is resonant on the $|0\rangle=|6S_{1/2} (F=4)\rangle$ $\rightarrow$ $|1\rangle=|6P_{3/2} (F=5)\rangle$ transition (Rabi frequency $\Omega_{P}=\Omega_{10}$), and two 510-nm EIT coupler-laser modes simultaneously drive the $|1 \rangle$ $\rightarrow$ $\vert 4\rangle=\vert 90S \rangle$ and
$|1 \rangle$ $\rightarrow$ $\vert 2\rangle=\vert 91S \rangle$ Rydberg transitions (respective Rabi frequencies $\Omega_{C^-} = \Omega_{41}$ and $\Omega_{C^+}=\Omega_{21}$). The Rydberg states $|4 \rangle$ and $|2 \rangle$ are near-resonantly coupled by the (same) RF field to Rydberg state $\vert 3\rangle=\vert 90P_{3/2}\rangle$, with respective Rabi frequencies $\Omega_{34} \approx \Omega_{32} =: \Omega_{RF}$.

In interferometric loop I, shown in Fig.~2a, the pair of two-photon couplings $|1\rangle\rightarrow|3\rangle$ are both exactly two-photon-resonant, with two-photon Rabi frequencies $\Omega_{A}$ and $\Omega_{B}$ as defined in Eq.~\ref{eq:int1}. The intermediate detunings from levels $|4\rangle$ and $|2\rangle$, labeled $\Delta_{90S}$ and $\Delta_{91S}$ in Fig.~2(a), are generally different but typically have similar magnitude. This system exhibits interference and phase sensitivity as exhibited in Eqs.~1 and~2.

In interferometric loop II, shown in Fig.~2b, the optical couplings $|1\rangle\rightarrow|2\rangle$ and $|1\rangle\rightarrow|4\rangle$ are both essentially on-resonance.
The RF field couples levels $|4\rangle$ and $|2\rangle$ in second order via level $|3\rangle$, with intermediate-state detuning denoted $\Delta_{90P}$. The detuning from two-photon resonance, $\Delta_{RF} = 2 \omega_{RF} - (W_2-W_4)/\hbar$ with level energies $W_i$, is typically held near zero, except in Fig.~4b where we explore the effects of a non-zero $\Delta_{RF}$. The interference in loop-II is seen by first noting that the two-photon RF coupling generates a pair of Autler-Townes~(AT)-split states, with AT splitting $\Delta_{AT}$, that are orthogonal superpositions of $|2\rangle$ and $|4\rangle$. These states are excited by the two optical coupler-laser modes with the Rabi frequencies $\Omega_{C^{\pm}}$. The quantum coherence of the bare atomic constituents $|2\rangle$ and $|4\rangle$ within the AT states, and the phase coherence of the two optical excitation paths with each other and with the RF field impinging onto the atoms, lead to an interferometric dependence of the net excitation amplitude on $\phi_{opt}-\phi_{RF}$. The signal in Loop II, observed as a function of the (unmodulated) coupler-laser detuning $\Delta_C$ and $\phi_{opt}$, consists of two bands that are separated in $\Delta_C$ by a two-photon AT-splitting $\Delta_{AT}$ (see Fig.~2d). Due to the orthogonality of the AT states, the EIT signals on the AT-pair of Loop-II bands have orthogonal dependencies on the phase $\phi_{opt}$.

The EIT probe transmission signal is related to the coupler-light-induced coupling strength $\vert \Omega \vert^2$, with $\Omega$ in the form of Eq.~2. Both Loops I and II allow measurement of an unknown $\phi_{RF}$ via tuning of $\phi_{opt}$, or vice versa. Loop II further yields
the RF field intensity by measuring the two-photon AT splitting $\Delta_{AT}$ and the detuning $\Delta_{90P}$.
In the case $\Delta_{RF}=0$, the RF intensity, $I_{RF}= (1/2) c \epsilon_0 E_{RF}^2$ with
RF electric field $E_{RF}$, follows from
$\Delta_{AT} = \Omega_{34} \Omega_{32} / (2 \Delta_{90P}) = d_{34} d_{32} E_{RF}^2 / (2 \hbar^2 \Delta_{90P}) $, with transition electric-dipole matrix elements $d_{nm}$.

Our experimental setup, illustrated in Fig.~\ref{fig:2}c, utilizes probe (852~nm) and coupler (509~nm) laser sources that are frequency-stabilized to $\sim$100~kHz. The 509~nm coupler laser beam is passed through an electro-optic phase modulator (PM) that is modulated at the RF frequency $\omega_{RF} \approx 5$~GHz. The first-order PM sidebands provide
a pair of coupler modes separated by $2 \omega_{RF}$ in frequency. The EIT coupler modes co-propagate and are counter-aligned with the EIT probe. The optical phase shift $\phi_{opt}$ is implemented
by phase-shifting the RF supplied by the signal generator to the PM. The optical field modes generating the couplings $\Omega_{C^\pm}$ transmit the phase $\phi_{opt}$ to the atoms, with opposite sign (see Eq.~(1)).
The signal generator also feeds a horn antenna that transmits the 5~GHz free-space signal wave to the cesium atoms, which are located in the far-field of the antenna. The EIT probe transmission signal is collected on a photoreceiver for read out of the RF interferometer while the frequency offset of the coupler-laser source, $\Delta_C$, is scanned. In the measurements we map the EIT probe transmission, $T$, in the ($\Delta_C$, $\phi_{opt}$) plane, revealing the RF phase $\phi_{RF}$ (Loops I and II) and intensity (Loop II).

\begin{figure}[h]
\includegraphics[width=8.5cm]{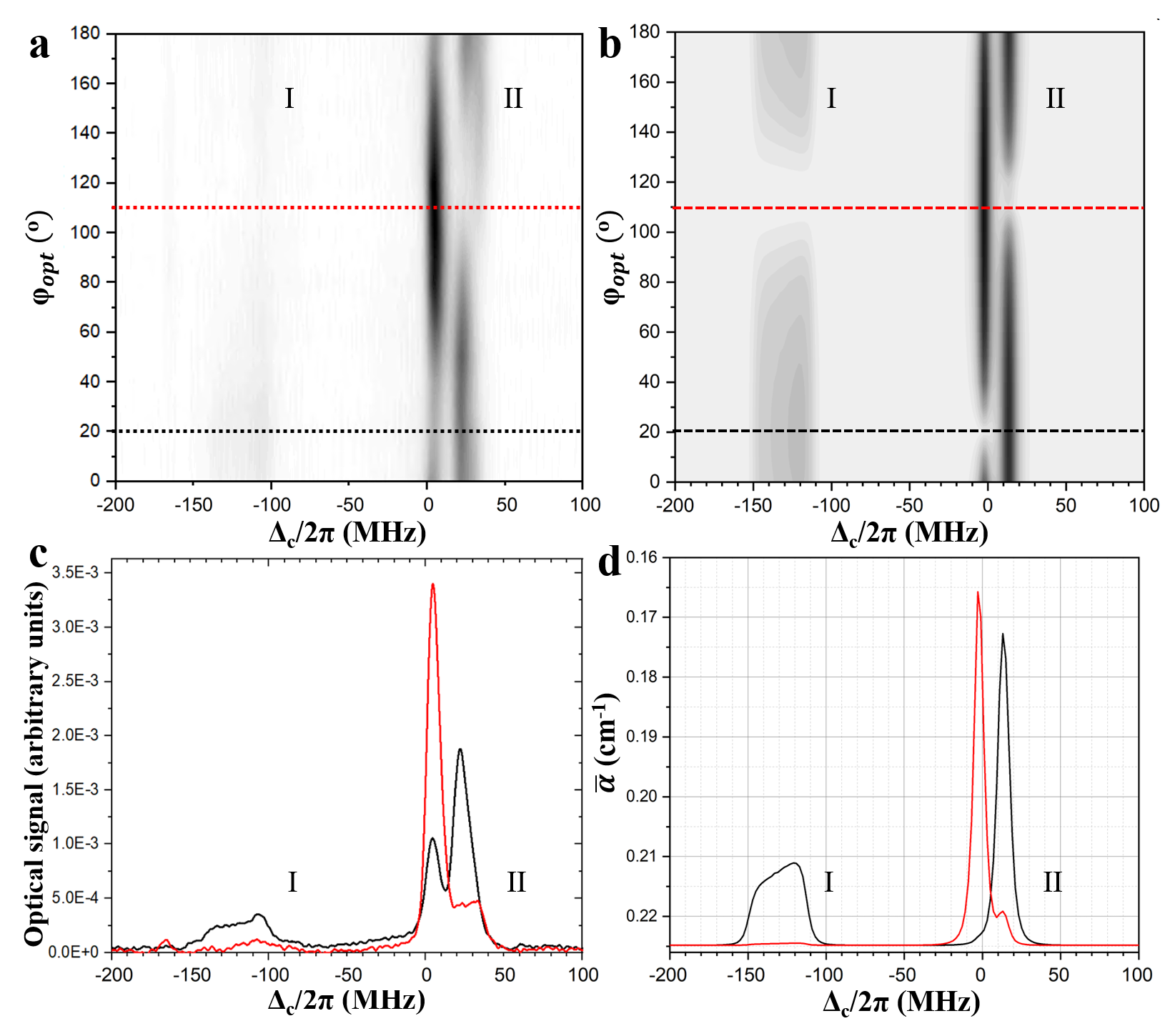}
\caption{(a) Measurement of $T(\Delta_C, \phi_{opt})$ to find the phase of a 5.092~GHz free-space RF wave. Signals from Loops I and II are labeled. (b) Calculation of cell probe absorption $\bar{\alpha}(\Delta_C, \phi_{opt})$
for conditions as in the experiment (a).  Experimental (c) and calculated (d) EIT spectra
$T(\Delta_C)$ and $\bar{\alpha}(\Delta_C)$ at fixed values of $\phi_{opt}$ indicated by dashed lines in (a) and (b), respectively.}
\label{fig:3}
\end{figure}

In Fig.~\ref{fig:2}d we determine a fixed, unknown phase $\phi_{RF}$ using Loop II by measuring
$T(\Delta_C, \phi_{opt})$ for $\omega_{RF}= 2 \pi \times 5.092$~GHz, a case that is near two-photon RF resonance $\Delta_{RF} \sim 0$ for the Rydberg states chosen in this work. Two EIT bands, corresponding to AT-split states separated by $\Delta_{AT} \approx 2 \pi \times 10$~MHz in $\Delta_C$, extend along the $\phi_{opt}$-direction and are modulated in $\phi_{opt}$ with a period of $\pi$, as generally expected by taking the square of Eq.~2. The two EIT bands of Loop II are shifted in $\phi_{opt}$ by $\pi/2$ relative to each other due to the orthogonality of the two AT-split states. The fringe visibility well exceeds 50\%, as seen more clearly in Fig.~3, with limiting factors discussed in our model below. We choose the inflection points of the Loop-II signals to assign a value to the phase $\phi_{RF}$ of the RF wave to be measured. In Fig.~2d, the measured RF phase is $\phi_{RF} = 85^\circ \pm 5^\circ$.

In Fig.~3a we show a similar measurement of both Loops I and II, and in Fig.~\ref{fig:3}c we plot individual EIT spectra $T(\Delta_C)$ at $\phi_{opt}$-values where the loop signals are extremal as a function of $\phi_{opt}$. There is a total of three EIT bands at certain values of $\Delta_C$, with the lowest-frequency band corresponding to Loop I and the two higher-frequency ones to the two AT-split bands of Loop II. Here, $\Delta_{AT} / (2 \pi) = 18 \pm 1$~MHz, and $\Delta_{90P} / (2 \pi) = - 130 \pm 15$~MHz, taken to be the separation between the center of the Loop-I band and the midpoint between the two Loop-II bands. The Loop-I signal is considerably weaker than that of Loop II due to the presence of a weak DC electric field.  The DC field shifts the Loop-I signal from its zero-field value of $\Delta_C / (2 \pi) = -55$~MHz to $\approx -130$~MHz, and broadens it over a range of about 30~MHz. Using the experimental values
for $\Delta_{AT}$ and $\Delta_{90P}$ and calculated values of $d_{34}=3942~ea_0$ and $d_{32}=3849~ea_0$, one finds $E_{RF} = 1.4 \pm 0.2$~V/m.

To numerically model the atom RF interferometer measurements in Figs.~\ref{fig:3}a and c, we obtain the steady-state density operator $\hat{\rho}$ of the Lindblad equation for the 5-level system of Fig.~\ref{fig:2}. The absorption coefficient of the cesium vapor for the probe beam is given by
\begin{equation}
\alpha  = \frac{2 \pi }{\lambda_P} \frac{2 n_V d_{10}}{\epsilon_0 E_P} \int P(v) {\rm{Im}} (\rho_{01}) dv
\end{equation}
Here, $n_V$ is the density of cesium atoms in the $F=4$ ground state at 293~K, $d_{10} = 1.9$~ea$_0$ is the probe electric-dipole matrix element, $\lambda_P$ is the probe-laser wavelength and $E_P$ is its electric-field amplitude, and $P(v)$ is the normalized one-dimensional Maxwell velocity distribution at 293~K. The coherence $\rho_{01}$ depends on Rabi frequencies, field phases, decay rates, atom-field detunings, and atom velocity $v$. 
The DC-field-free dressed Rydberg-level detunings at $v=0$ are
$\Delta_{90S}=2\pi \times 55.1~$MHz$ - \Delta_C $,
$\Delta_{90P}= - \Delta_C $, and
$\Delta_{91S}=2\pi \times 55.1~$MHz$ - \Delta_C $.
We use an optical EIT probe Rabi frequency at the beam center of $\Omega_{10} = 2 \pi \times 7.7$~MHz, and EIT coupler Rabi frequencies at the beam center of $\Omega_{21} \approx \Omega_{41} = 2 \pi \times 1.0$~MHz. In Fig.~\ref{fig:3}a and~\ref{fig:3}c the measured RF Rabi frequencies are $\Omega_{34}=64~$MHz and $\Omega_{23}=62~$MHz.

\begin{figure}[htb]
\includegraphics[width=8cm]{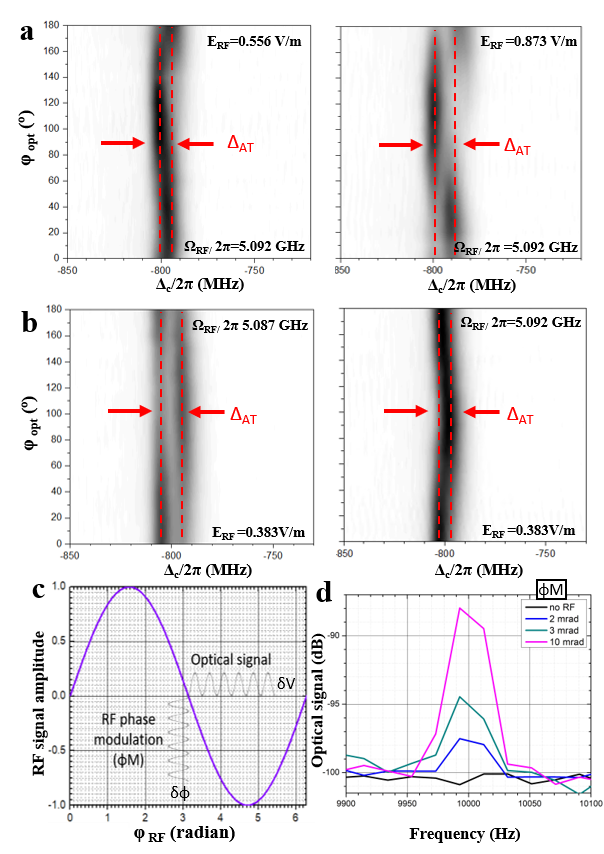}
\caption{(a) Loop-II signal for the indicated RF electric fields, $E_{RF}$, for on-resonant two-photon RF transition ($\Delta_{RF} \approx 0$). (b) Loop-II signal for two-photon detuning $\Delta_{RF} \approx 0$ (right panel)
and $\Delta_{RF} \approx 10$~MHz (left panel), for fixed and small $E_{RF}$. (c) Mapping of phase modulation (PM), $\delta \phi$, into a modulated EIT signal, $\delta V$. (d) Measured power spectrum of the EIT signal, $\delta V$, for the indicated values of $\delta \phi$ for a PM frequency of 10~kHz.}
\label{fig:4}
\end{figure}

We compute weighted averages of $\alpha$ from Eq.~3 over the transverse Gaussian profile of the laser beams
and over distributions of the DC electric field, $P_E(E)$, leads to the shift and broadening of the Loop-I signal in Figs.~3a and c.  Here, the measured results are reproduced by a flat distribution $P_E(E)$ ranging from $E$=29~mV/cm to 41~mV/cm. Figures~\ref{fig:3}b and d show the calculated averaged absorption coefficient, $\bar{\alpha}(\Delta_C, \phi_{opt})$, and respective cuts at selected phases $\phi_{opt}$, for conditions as in Figs~\ref{fig:3}a and c.  Good agreement is found between experiment and calculation, reproducing the relative phases, AT splittings and linewidths of the signal bands.

The higher-frequency Loop-II band in the experimental data $T(\Delta_C, \phi_{opt})$ exhibits a correlation between the phase of the EIT signal and the detuning $\Delta_C$. This is attributed to a spatial correlation between the DC field and the optical-beam propagation phase $\Delta k \, x$, with $\Delta k$ denoting the wavenumber-difference between the two coupler modes and $x$ the position along the coupler-beam direction. The reduced
visibility in the phase dependencies of the Loop-I and II signals in Fig.~3c relative to those in Fig.~3d is attributed to the propagation phase $\Delta k \, x$.

In Fig.~\ref{fig:4}a we verify the dependence of the Loop-II splitting, $\Delta_{AT}$, on $E_{RF}$ for the case of two-photon RF resonance, $\Delta_{RF} \approx 0$. In that case, one expects $\Delta_{AT} \propto E_{RF}^2$, in good qualitative agreement with the measurements shown in Fig.~4a.
For non-zero $\Delta_{RF}$, the Loop-II splitting is given by $\sqrt{\Delta_{AT}^2 + \Delta_{RF}^2}$, the effective, off-resonant two-photon Rabi frequency. This behavior is evident in Fig.~4b, where $\Delta_{RF} \approx 10$~MHz in the left and $\Delta_{RF} \approx 0$ in the right panel. The visibility of the phase-dependent modulation of the Loop-I and Loop-II bands is expected to diminish with increasing $\vert \Delta_{RF} \vert$, which is supported by the measurements in Fig.~4b. The measurements in Figs.~4a and~4b agree well with simulations (not shown).

The static RF phase resolution in the phase measurements in Figs.~2 and~3 is about $5^\circ$. In the last component of the present work, we determine the phase resolution achievable using a dynamic interferometric method. For a fixed optical phase near the inflection point $\frac{\partial^2}{\partial \phi_{opt}^2} T(\Delta_{C0}, \phi_{opt}) =0$, with $\Delta_{C0}$ set at a Loop II-band, we apply a weak phase modulation (PM) of known amplitude $\delta \phi_{RF}$ to the RF field, with a PM frequency of 10~kHz. Since variations of optical and RF phase have equivalent effects on the optical signal $T$, lock-in detection of $T$ at the PM frequency yields a modulation amplitude $\delta T = \left[ \frac{\partial}{\partial \phi_{opt}} T(\Delta_{C0}, \phi_{opt}) \right] \delta \phi_{RF}$. In Figs. 4c and~d we observe a minimum detectable $\delta \phi_{RF}\approx 0.1^\circ = 2$~mrad, limited by the RF hardware.

In summary, we have realized an atom RF interferometer, employed it for all-optical phase measurement of an RF wave with a cesium vapor sensor, and developed a quantum model that reproduces all observed features.  A phase resolution of 2~mrad was achieved and RF phase and field measurements were demonstrated at sub-millimeter optical spatial resolution with the atomic sensor.  Atom RF interferometry provides a platform to enable new capabilities in optical and RF phase, frequency, and amplitude signal detection, measurement, and imaging, relevant to a broad range of applications in RF and communications technology.

This work was supported by Rydberg Technologies Inc.


%

\end{document}